\documentclass{article}
\usepackage[utf8]{inputenc}
\usepackage[T1]{fontenc}
\usepackage{bm}
\usepackage{mathtools}
\usepackage{amssymb}
\usepackage{subcaption}
\usepackage{float}
\usepackage{lineno}
\usepackage{setspace}
\usepackage{authblk}

\title{On the role of solute drag in reconciling laboratory and natural constraints on olivine grain growth kinetics}
\author[1,2]{J. Furstoss}
\author[1]{C. Petit}
\author[3]{A. Tommasi}
\author[1]{C. Ganino}
\author[2]{D. Pino Mu\~{n}oz}
\author[2]{M. Bernacki}
\affil[1]{Universit\'e Nice C\^ote d'Azur, CNRS, OCA, IRD, G\'eozur, France}
\affil[2]{MINES ParisTech, PSL Research University, CEMEF-Centre de mise en forme des mat\'eriaux, CNRS UMR 7635, France}
\affil[3]{G\'eosciences Montpellier, Univ. Montpellier, CNRS, Montpellier, France}

\usepackage{graphicx}

\begin{document}

\maketitle

\section*{Abstract}

We investigate the effect of solute drag on grain growth (GG) kinetics in olivine-rich rocks through full field and mean field modelling. Considering a drag force exerted by impurities on grain boundary migration allows reconciling laboratory and natural constraints on olivine GG kinetics. Solute drag is implemented in a full field level-set framework and on a mean field model, which explicitly accounts for a grain size distribution. After calibration of the mean field model on full field results, both models are able to both reproduce laboratory GG kinetics and to predict grain sizes consistent with observations in peridotite xenoliths from different geological contexts.



\section{Introduction}

Olivine is the major constituent of the Earth upper mantle, and its grain growth (GG) kinetics is of major importance in several geodynamic processes. In fact, a variation in the mean grain size of a mantle rock may drastically change its mechanical behavior through the grain size dependence of the diffusion creep regime \cite{karato1986rheology}. In this regime, grain size reduction produces significant weakening, whereas grain growth may lead the rocks into the grain-size independent, but stress-dependent dislocation creep regime. A variation in the mean grain size of olivine rocks may therefore produce marked changes in the upper mantle rheology, which may control strain localization \cite{braun1999simple}. Consistently, preservation of small grain sizes (mylonitic or ultramylonitic microstructures within ductile shear zones \cite{vissers1991shear}) has been proposed as one of the mechanisms allowing to preserve weak plate boundaries over geological times \cite{BercoRicardNature}.

However, numerical models \cite{Furstoss2018, chu2012olivine} based on experimental data on olivine GG \cite{Karato89, Ohuchi2007, Hiraga2010} cannot explain the persistence of small grain sizes through million years in pure olivine rocks. Although they simulate well the experimental data, these models also fail to predict the commonly observed plurimillimetric grain sizes observed in dunites when run on geologically relevant timescales (i.e., over millions of years), for which they predict meter scale grain sizes \cite{Furstoss2018, chu2012olivine}. To obtain a GG kinetics compatible with natural observations in upper mantle rocks, the presence of second phases is often considered \cite{Hiraga2010, Furstoss2020}. Nevertheless, to preserve small olivine grain sizes over millions of years, these models have to introduce some questionable features, such as very slow second phase GG \cite{NakakojiHiraga2018Part2}, an enhanced mixing between phases \cite{BercoRicard1} or the presence of small fixed particles \cite{Furstoss2020} to increase the impediment of olivine grain boundary migration (GBM).

Our understanding of olivine GG in itself is limited by the inconsistency between laboratory and natural time and spatial scales. All GG models are calibrated on experimental data obtained for ultra-fine grains ($1$ to $30 \mu m$), for which the grain size evolution is detectable at the experimental timescale (a few hours to several days), and then extrapolated to natural conditions. As a consequence, if a physical mechanism becomes prominent only for grain sizes larger than $50 \mu m$, its effect will not be captured in the experiments. Such a physical mechanism may be the drag force exerted by impurities also called solute drag, which has been very seldom considered in studying GG in rocks (orthopyroxene \cite{PyroxeneGG}, halite \cite{guillope1979dynamic}) and never accounted for in modeling GG of olivine.

In the present work we tested the effect of solute drag on the GG kinetics of olivine. We performed 2D full field GG simulations using a level-set (LS) framework \cite{bernacki2011level} in which we implemented the solute drag effect. These models show that considering this mechanism allows for consistent simulation of olivine GG kinetics at both experimental and upper mantle conditions. Finally we adjust a mean-field model on results of the full field simulations to propose an analytical expression allowing to compute efficiently the grain size evolution within geodynamic large scale models accounting for grain size-dependent rheologies.

\section{Modeling solute drag}
\label{SoluteDragTh}

\begin{figure}[H]
	\begin{center}
		\hbox to \hsize{\hss\includegraphics*[width=\textwidth]{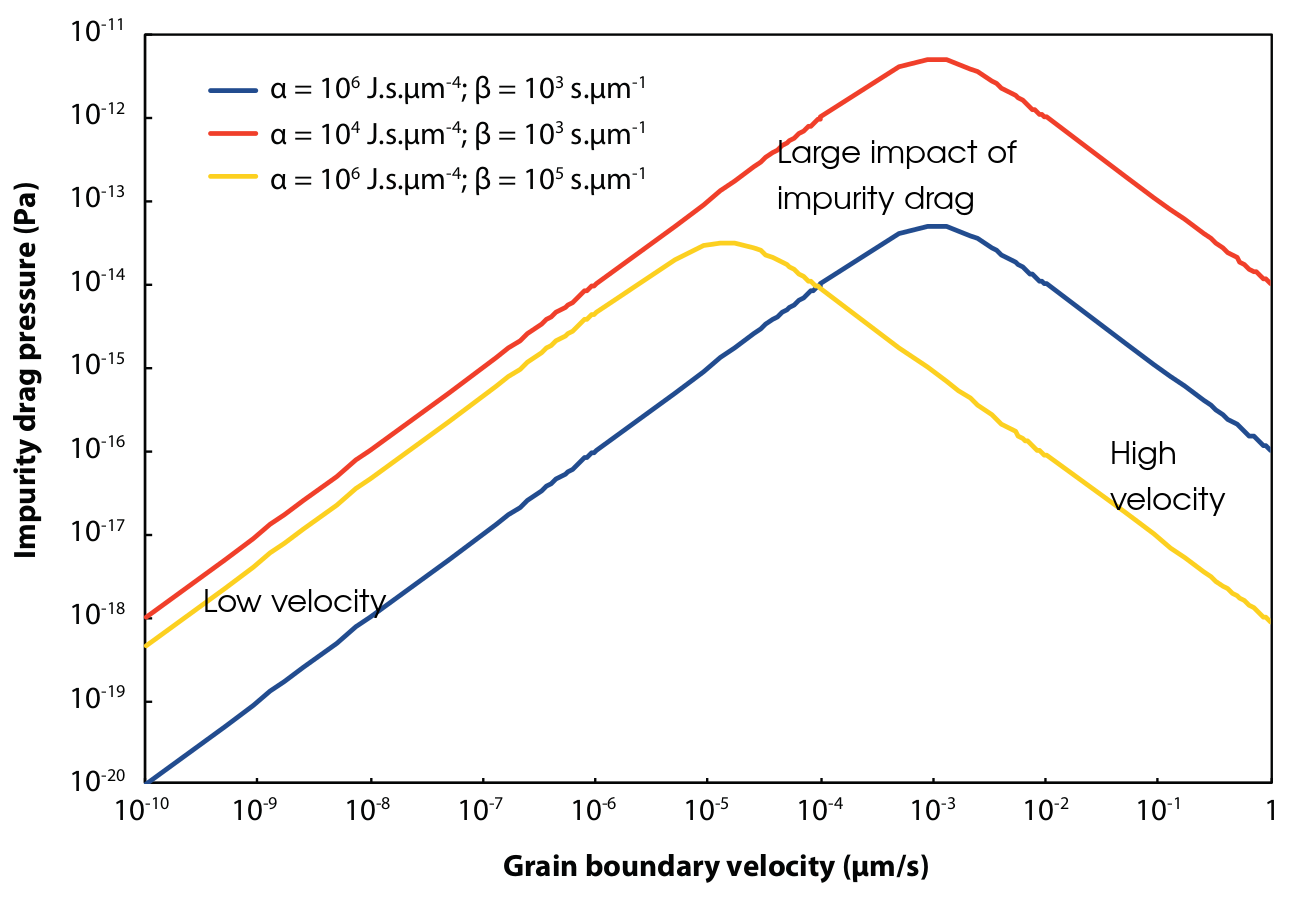}\hss}
		\caption{Solute drag pressure intensity as a function of the grain boundary velocity calculated using eq.\ref{EqImpurityDragPressure} for $c_0 = 100$ppm and different values of $\alpha$ and $\beta$.}
		\label{fig0}
	\end{center}
\end{figure}

Impurities present within the crystal lattice and segregated at grain boundaries can have an impact on their migration kinetics through the so called solute drag effect. The grain boundary migration velocity ($v$) is generally expressed by \cite{Humphreys} :

\begin{equation}
v = M P,
\label{eqGBM}
\end{equation}

\noindent
where $M (m^4.J^{-1}.s^{-1})$ is the grain boundary mobility and $P$ is the sum of the pressures exerted on the grain boundary. The solute drag effect can be described in terms of dragging pressure exerted on the grain boundary, which is a function of the grain boundary velocity and impurities concentration and nature. The quantification of this drag pressure has been studied theoretically \cite{lucke1957quantitative} . Its intensity follows three main regimes (high, intermediate and low velocity), similarly to well known dislocation impediment by Cottrell atmospheres \cite{cottrell1949dislocation}. In fact, impurities segregated around grain boundary interact with it and when the grain boundary migrates the impurity cloud tends to accompany it. In the high velocity regime, the grain boundary moves so fast that the segregated impurities cannot follow the interface and the interactions between the impurities and the grain boundary are much reduced. In the low velocity regime, the impurity cloud can stay segregated around the grain boundary, but the intrinsic drag of the grain boundary (due to drag exerted by the intrinsic defects within the interface) is generally higher than the drag exerted by the impurities (Fig.\ref{fig0}). Between these two velocities, a third regime exists where the impact of solute drag on GBM is the most important.

\noindent
Even if the drag effect is expected to follow different mathematical relationships depending on velocity regime \cite{lucke1957quantitative}, an unified expression for the drag pressure $P_i$ exerted by a $c_0$ concentration of impurities (within the grain matrix) describing the two velocity regimes has been proposed by \cite{cahn1962impurity} :

\begin{equation}
P_i = \frac {\alpha v c_0} { 1 + \beta^2 v^2 },
\label{EqImpurityDragPressure}
\end{equation}

where $v$ is the grain boundary velocity, $c_0$ the concentration of impurities within the bulk, $\alpha (J.s.m^{-4})$ and $\beta (s.m^{-1})$ are two parameters modulating the intensity of the drag. $\alpha$ controls the intensity of the solute drag pressure and $\beta$ constrains the grain boundary velocity which is the most impacted by the presence of the impurities (Fig.\ref{fig0}). The drag effect is most effective for grain boundary velocities close to of $1/\beta$.

\noindent
The $\alpha$ and $\beta$ parameters can be expressed as integrals of functions $D(x)$ and $E(x)$ representing the variation of the impurity diffusion coefficient and interaction energy, respectively, along the grain boundary normal ($x$). As these two functions are difficult to constrain experimentally, their mathematical expressions are generally hypothesized in an integrable way, which permits to obtain analytical expressions for $\alpha$ and $\beta$ \cite{cahn1962impurity} :

\begin{equation}
\alpha = \frac { N_v (k_b T)^2 }{ E_0 D } ( sinh(\frac {E_0}{k_b T}) - \frac {E_0}{k_b T} ),
\label{eqAlpha}
\end{equation}

and,

\begin{equation}
\beta ^2 = \frac { \alpha k_b T \delta }{ 2 N_v E_0 ^2 D },
\label{eqBeta}
\end{equation}

where $N_v (m^{-3})$ is the number of atoms per unit volum, $k_b$ the Boltzmann constant, $E_0 (J)$ the interaction energy, $D (m^2 .s^{-1})$ the diffusion coefficient of the impurity within the grain matrix and $\delta (m)$ is the characteristic segregation length of the impurities around the grain boundary.

\subsection{Semi-explicit implementation of solute drag within the level-set framework}
\label{LSImpl}

The LS framework \cite{CemefDyn}, already used to model olivine GG \cite{Furstoss2018, Furstoss2020}, proposes an implicit description of the polycrystal through the use of LS functions representing the signed distance function to the grain boundaries surrounding the grain they represent (positive inside the grain and negative elsewhere) in a finite element (FE) context. The microstructural evolution is simulated by moving LS functions according to physical laws describing GBM \cite{CemefGG2D} or by creating LS functions to represent new grains nucleated during the recrystallization processes \cite{DRXLudo}. Theoretically, each grain of a polycrystal is represented by its own LS function. In order to reduce the computation time and memory storage, several non-neighboring grains in the initial microstructure can be grouped to form Global Level Set (GLS) functions thanks to a graph coloration technique. Re-coloration technique is used to avoid numerical grain coalescence during grain boundary motion \cite{lvlsetcont, CemefStatic}. The GLS functions displacement is computed within an efficient FE \cite{lvlsetcont, lvlsetreini} framework using anisotropic mesh refinement around interfaces \cite{remesh}. The initial microstructure is generated using a Vorono\"\i-Laguerre Dense Sphere Packing algorithm \cite{hitti2012precise} allowing to respect precisely an imposed initial grain size distribution.

\noindent
If GG is solely controlled by capillarity (reduction in the grain boundary energy) and solute drag, the grain boundary velocity can be expressed as the sum of the capillarity pressure ($P_{capi}$) and impurity drag pressure ($P_i$ described by Eq.\ref{EqImpurityDragPressure}) as :

\begin{equation}
\vec{v} = M(P_{capi} - P_i)\vec{n} = M ( - \gamma \kappa - \frac{\alpha c_0 \vec{v} \cdot \vec{n} }{1 + \beta^2 v^2} ) \vec{n},
\label{EqVImpur}
\end{equation}

\noindent
where $v^2 = ||\vec{v}||^2$, $M (m^4.J^{-1}.s^{-1})$ and $\gamma (J.m^{-2})$ are the grain boundary mobility and energy respectively, $\kappa$ is the grain boundary local curvature (in 2D) or the sum of the main local curvatures (in 3D) and $\vec{n}$ the outward unit normal to the grain boundary. 

\noindent
By introducing the GLS function $\Phi_i$ to represent the grains $i$, we can do the following implicit and explicit first order time discretizations, for the temporal derivative of $\Phi_i$ :

\begin{equation}
\frac{\partial \Phi_i }{\partial t} = \frac{\Phi^{t+\Delta t}- \Phi^t}{\Delta t},
\label{DiscretisationTemporelle_1}
\end{equation}

\noindent
for the GLS function velocity : 

\begin{equation}
\vec{v} = \frac{\Phi^{t+\Delta t}- \Phi^t}{\Delta t} \vec{n} = - \frac{\Phi^{t+\Delta t}- \Phi^t}{\Delta t} \vec{\nabla}\Phi,
\label{DiscretisationTemporelle_2}
\end{equation}

\noindent
and for the velocity squared : 

\begin{equation}
\vec{v}^2 = (\frac{\Phi^{t}- \Phi^{t-\Delta t}}{\Delta t})^2 = v_{old}^2.
\label{DiscretisationTemporelle_3}
\end{equation}

\noindent
The displacement of the GLS functions are then computed using the convective LS equation \cite{osher1988fronts} :

\begin{equation}
\frac{\partial \Phi_i }{\partial t} + \vec{v} \cdot \vec{\nabla}\Phi_i = 0.
\label{eqConvLS}
\end{equation}

\noindent
Considering the geometrical properties of LS function $\kappa = - \Delta \Phi$, $\vec{\nabla}\Phi.\vec{\nabla}\Phi = 1$, $\vec{\nabla}\Phi = - \vec{n}$, using the above time discretization and substituting Eq.\ref{EqVImpur} within the above convective LS equation, we obtain the FE strong formulation :

\begin{equation}
\mathcal{M}\frac{\Phi^{t+\Delta t}}{\Delta t} - M\gamma \Delta \Phi^{t+\Delta t} = \mathcal{M}\frac{\Phi ^t}{\Delta t},
\label{FinalStrongForm}
\end{equation}

\noindent
where :

\begin{equation}
\mathcal{M} = 1 + \frac{M\alpha c_0 }{1+\beta^2 v_{old}^2}.
\label{eqMassTerm}
\end{equation}

\noindent
The fact that Eq.\ref{eqMassTerm} uses $v_{old}^2$ instead of $v^2$ is a simplification that reduces the non-linear behaviour of the problem. In practical terms, the time marching scheme uses a small timestep, thus the error of replacing $v^2$ by $v_{old}^2$ is small. However it greatly simplify the numerical scheme allowing its implementation on a generic FE code. This formulation is similar to the diffusion formulation classically used for capillarity driven GG in the LS formalism \cite{CemefGG2D}. By analogy with the heat equation, $\mathcal{M}$ is equivalent to a mass term (i.e. the product between the specific heat capacity and density). The main difference with the classic LS approach for capillarity-driven GG when solute drag is modelled is that the mass term differs from one and depends on the velocity of the LS function at time $t - \Delta t$. This heterogeneous mass term is computed on each node of the mesh and linearly interpolated.

\noindent
A major drawback of the LS approach lies in the fact that during grain boundary migration, the GLS are no longer distance functions $||\vec{\nabla}\phi|| \neq 1$. This is particularly problematic when a remeshing technique depending on the distance property is used at grain interfaces. In addition, the new diffusive formulation proposed in Eq.\ref{FinalStrongForm} requires a distance function as it is based on the respect of $||\vec{\nabla}\phi|| = 1$, at least in a thin layer around the interface. For these reasons, the GLS functions need to be reinitialized at each time step in order to restore their metric property. Numerous approaches exist for this reinitialization procedure. Here we use a new direct fast and accurate approach usable in unstructured FE mesh proposed  by Shakoor et al. \cite{lvlsetreini}. The residual errors inherent to this approach are discussed in \cite{florez2020new}.

\noindent
The introduction of the solute drag pressure is expected to reduce the GG kinetics. Thus its influence has to be accounted for in the adaptative time stepping scheme in order to allow larger steps when the grain size evolves slowly. The timestep is computed by imposing a maximal incremental displacement corresponding to a given fraction ($H$) of the LS reinitialized width ($E_p$) :

\begin{equation}
\Delta t = \frac{H E_p}{v_m},
\label{CalculDt}
\end{equation}

\noindent
where $v_m$ is the grain boundary mean velocity :

\begin{equation}
v_m = M ( \frac{\gamma}{\bar{R}} - \frac{c_0 \alpha | \bar{\dot{R}} | }{ 1 + \beta ^2 \bar{\dot{R}} ^2} ),
\label{Calculvm}
\end{equation}

\noindent
where $\bar{R}$ and $\bar{\dot{R}}$ are the mean grain radius and its temporal evolution respectively. In order to exclude a negative timestep value, we impose a lower bound for $v_m$ to $M\gamma / R_{max}$ where $R_{max}$ is the maximum grain radius.

\begin{figure}[H]
	\begin{center}
		\hbox to \hsize{\hss\includegraphics*[width=\textwidth]{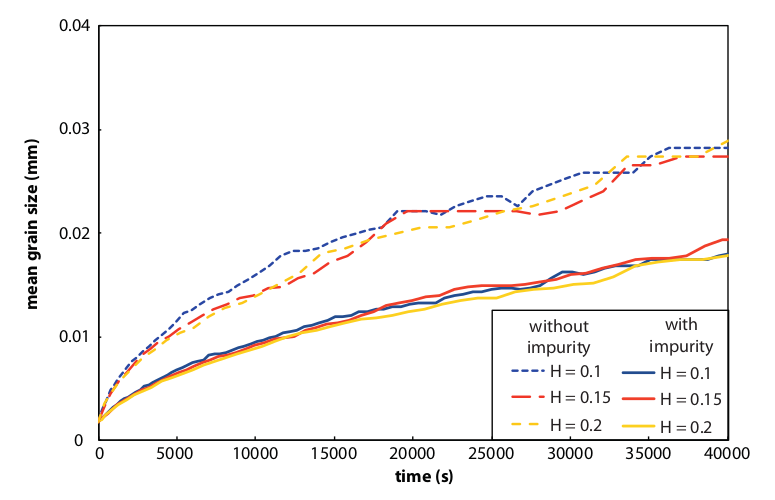}\hss}
		\caption{Full field model of the mean grain radius evolution using $M = 2.9\cdot 10^{-2} mm^4.J^{-1}.s^{-1}$, $\gamma = 10^{-6} J.mm^{-2}$ for a case without impurities (dashed lines) and for a case with $c_0 = 1000 ppm$, $\alpha = 10^5 J.s.mm^{-4}$ and $\beta = 10^5 s.mm^{-1}$ (solid lines). The $H$ coefficient controls the timestep (Eq.\ref{CalculDt}).}
		\label{fig1}
	\end{center}
\end{figure}

The explicit time discretization in Eq.\ref{DiscretisationTemporelle_3} may have an impact on the numerical resolution if the non-linearity of the grain boundary velocity (Eq.\ref{EqVImpur}) is strong. This has been evaluated by performing computations with different timesteps (Fig.\ref{fig1}), which show that mean grain size evolution does not depend on $H$, neither for the reference case without solute drag ($\mathcal{M} = 1$), nor for the case with solute drag.

\noindent
The GG kinetics when solute drag accounts for impurities exhibit the expected trend, being slower than in cases without drag (Fig.\ref{fig1}). Thus, the strong formulation (Eq.\ref{FinalStrongForm}) for the displacement of LS functions accounting for solute drag can be validated.

\subsection{Mean field approach for GG with solute drag}
\label{MeanField}

To construct a mean field model describing GG kinetics accounting for solute drag, it is important to account for the initial grain size distribution. In fact, computing GG kinetics accounting only for mean grain size evolution will hide the dispersion of the solute drag pressures exerted within the microstructure, due to variations in the capillarity force, which is a function of grain size. This will result in three clear stages of grain size evolution, an initial mean grain size evolution unimpacted by solute drag (high velocity regime), a second phase, the most impacted by solute drag, where mean grain size will be quasi-static and a final phase also unimpacted by solute drag (low velocity regime). To account for the coexistence of these three phases within the microstructure and predict realistic mean grain size evolutions, the initial dispersion of individual grain sizes and their evolution rates have to be considered.

\noindent
For this purpose, we adapted the Hillert's model \cite{hillert1965theory}, which proposes a discrete representation of the grain size distribution (GSD) in the microstructure. By considering $R_i$ the radius of the i-grain size bin, the Hillert's model allows computing the evolution of each bin by accounting for the capillarity pressure expressed as $M \gamma ( \frac{ \overline{ R } } { \overline{R^2} } - \frac{ 1 } { R_i } )$ multiplied by a coefficient ($=\kappa \frac{ \overline{ R }^2 } { \overline{ R^2 } }$ with $\kappa \approx 1.6$, \cite{CemefGG3D}), which is adjusted based on experimental and/or full field results. This procedure enables to follow the evolution of each bin of the initial GSD. This mean field approach is able to reproduce, in terms of GSD, the predictions of full field simulations in context of pure GG (in 2D \cite{CemefGG2D} and in 3D \cite{CemefGG3D}) even for complex initial GSD (like bimodal ones).

\noindent
To account for solute drag, we subtract from the capillarity pressure the solute drag pressure (Eq.\ref{EqImpurityDragPressure}) replacing $v$ by $\dot{R_i}$ and $v^2$ by $\dot{R}_{i, old}^2$ (i.e. the grain size evolution rate at the precedent increment), which gives :

\begin{equation}
\dot{R_i} = \kappa \frac{ \overline{ R }^2 } { \overline{ R^2 } } M ( \gamma ( \frac{ \overline{ R } } { \overline{R^2} } - \frac{ 1 } { R_i } ) - \frac { \alpha c_0 \dot{R_i} } { 1 + \beta ^2 \dot{R}_{i, old}^2 } ),
\label{HillertModified}
\end{equation}

\noindent
which can be reformulated as :

\begin{equation}
\dot{R_i} = \gamma ( \frac{ \overline{ R } } { \overline{R^2} } - \frac{ 1 } { R_i } ) / ( \frac { \overline{ R^2 } } { M \kappa \overline{ R }^2 } + \frac { \alpha c_0 } { 1 + \beta ^2 \dot{R}_{i, old}^2 } ),
\label{HillertModified_2}
\end{equation}

\noindent
To account for topological effects or non-uniform mass term along grain boundaries, equation \ref{HillertModified} is generalized through the following expression :

\begin{equation}
\dot{R_i} = \gamma ( \frac{ \overline{ R } } { \overline{R^2} } - \frac{ 1 } { R_i } ) / ( \frac { \overline{ R^2 } } { M \kappa \overline{ R }^2 } + \frac { C_{\alpha} \alpha c_0 \dot{R}_{i, old}^{n} } { 1 + ( C_\beta \beta ) ^2 \dot{R}_{i, old}^2 } ),
\label{HillertModifiedMF}
\end{equation}

\noindent
where $C_{\alpha}$, $C_\beta$ and $n$ are mean field parameters which have to be calibrated on full field simulations. It can be noticed that the equation \ref{HillertModifiedMF} is equivalent to the non-generalized form (Eq.\ref{HillertModified_2}) for $n=0$ and $C_{\alpha} = C_{\beta} = 1$.

\begin{figure}[H]
	\begin{center}
		\hbox to \hsize{\hss\includegraphics*[width=\textwidth]{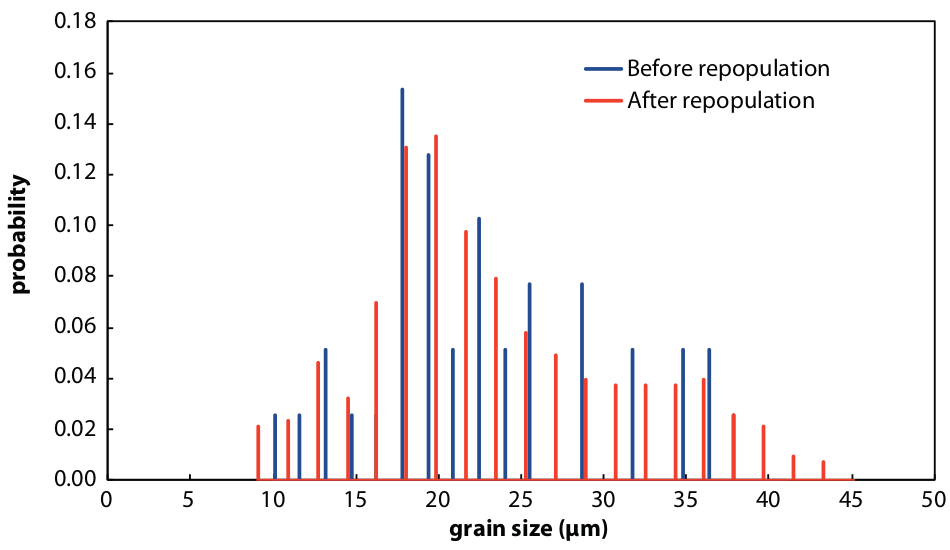}\hss}
		\caption{Example of GSD (in number) before and after a repopulation step with a minimal bin number of 40. Both histograms are constructed using 20 classes ranging between the radii of the bins with the minimal and maximal grain radius.}
		\label{figRepopDistrib}
	\end{center}
\end{figure}

\noindent
In practice, each bin contains an unique grain of radius $R_i$. We begin with a list of grain radii generated from an imposed initial GSD and compute iteratively their evolutions. If a grain radius becomes lower than $0.1 \mu m$ it is considered as consumed by the growth of neighboring grains and removed from the grain list. As this mean field model is intended to be used for long term calculations (Myr), the initial number of bins needed to preserve a representative number of grains after long annealing time is very large and the computational cost becomes prohibitive. Thus, we use a repopulation strategy allowing to do the calculation with a reasonable number of bins all along the simulation. To do so, when the number of bins is less than a minimal number, we repopulate the bin list by adding ten new bins for each existing bin $R_i$ with radii ranging between $0.9R_i$ and $1.1R_i$ with a step of $0.02R_i$. Those range of values have been chosen in order minimize the difference between GSD before and after repopulation step (see an example in Fig.\ref{figRepopDistrib}). The minimal number of bins is fixed at 40 based on a convergence study.

\section{Solute drag in olivine}\label{SoluteDragOl}

In mantle rocks, several incompatible elements and impurities are present in concentrations ranging from few to hundreds $ppm$ \cite{de2010trace}. Some of them are more or less homogeneously dispersed through the bulk material, but others can be enriched at grain boundaries \cite{suzuki1987seggregation}. The elements that can influence grain growth through solute drag are those that exhibit a partitioning between grain interiors and boundaries. In olivine-rich rocks, nickel (Ni), aluminum (Al) and calcium (Ca) \cite{suzuki1987seggregation, hiraga2003seggregation, hiraga2004seggregation} display a strong partitioning between olivine grain matrix and boundaries, the latter being often qualified as \guillemotleft incompatible element reservoirs \guillemotright. The major parameters controlling their effect on GBM are the element concentration, diffusion coefficient and interaction energy with grain boundary.
The latter will control at first order the intensity of the drag pressure exerted by the impurity. As interstitials, vacancies and even grain boundaries could have a non-null electrical charge in materials as olivine, the full formulation of this energy term should include, in addition to an elastic part, an electrostatic component. However for sake of simplicity we will assume a pure elastic interaction corresponding to the lattice distortion due to the misfit between the impurity size and the size of the typical host ion the impurity replaces. Close to and within olivine grain boundaries the impurities may replace Mg or Si ions (a depletion in Mg is observed at grain boundaries) \cite{hiraga2004seggregation}. Within the grain interiors, Mg ions are located in the octahedral M sites, and the interaction energy between impurity and grain boundary can be computed, as a first order approximation, using the characteristic length $r_0$ of those sites \cite{hiraga2004seggregation} :

\begin{equation}
E_0 = 4 \pi E_{\alpha} ( \frac{r_0}{2}( r_i - r_0 )^2 + \frac{1}{3}( r_i - r_0 )^3 ),
\label{EqInteractionEnergy}
\end{equation}

with $E_{\alpha}$ is the Young's modulus of the M lattice site and $r_i$ is ionic radius of the impurity. $E_\alpha$ is close to the bulk Young's modulus and $r_0$ can be computed from the length of the bonds between M and oxygen sites, and between oxygen and oxygen sites \cite{hiraga2004seggregation}. The temperature dependency of these parameters is small and we will consider in the following $E_\alpha = 159GPa$ and $r_0 = 0.064 nm$ \cite{hiraga2004seggregation}.

\noindent
Using the above expression and Eqs.\ref{EqImpurityDragPressure}, \ref{eqAlpha} and \ref{eqBeta} one can compute the drag pressure exerted by Ni ($r_i = 0.069 nm$), Al ($r_i = 0.054 nm$) and Ca ($r_i = 0.1 nm$) for different grain boundary velocities and for natural characteristic $c_0$ concentrations of $2500$, $50$ and $100 ppm$ respectively \cite{de2010trace}. Considering the diffusion coefficients of the three elements at $1573K$ \cite{spandler2007diffusion, coogan2005diffusion}, we estimate that the characteristic drag pressure exerted by Ca is at least 1 order of magnitude higher than the ones exerted by Ni and Al. Thus in the following, the only impurity considered for solute drag will be Ca.

\noindent
The Ca solute drag parameters $\alpha_{Ca}$ and $\beta_{Ca}$ can be computed using equations \ref{eqAlpha} and \ref{eqBeta} with a segregation length ($\delta$) of $5nm$ \cite{hiraga2004seggregation} and a calcium diffusion coefficient ($D$) taken as Arrhenius's laws where the reference value and activation energy are equal to $D_0 = 3.16.10^{-6} mm^2.s^{-1}$, $Q_D = 200kJ.mol^{-1}$ \cite{coogan2005diffusion} respectively. We obtain $\alpha_{Ca} = 3.10^{7}J.S.mm^{-4}$ and $\beta_{Ca} = 9.6.10^{5} s.mm^{-1}$ at 1573K and $\alpha_{Ca} = 3.10^{11}J.S.mm^{-4}$ and $\beta_{Ca} = 2.8.10^{9} s.mm^{-1}$ at 1073K.

\section{Laboratory and natural constraints on grain sizes in olivine-rich rocks}

The main goal of this study is to show that laboratory experiments and natural observation on olivine GG can be reconciled by accounting for Ca solute drag. In the following section, we define a priori constraints on GG kinetics provided by natural and experimental data.

\subsection{Laboratory constraints}

The annealing experiments for ultrafine grained natural San Carlos olivine \cite{Karato89} seem appropriate to define the experimental reference, particularly the dry runs at $1573K$. This sample has probably as much impurities as a natural mantle rock (because it was synthesized using crushed grains of natural olivine), so the solute drag model should fit these data. These experiments allow constraining grain growth kinetics for short times ($\leq 10h$) and small grain sizes ($2 - 25 \mu m$).

\subsection{Natural constraints}
\label{BiblioGeol}

Natural constraints on GG kinetics have to be considered cautiously because of the large number of uncertainties in the determination of physical conditions of natural systems (initial size distribution, precise thermal history, pure static conditions, etc.). Contraints on thermal history, strain and grain size evolution of peridotite xenoliths can be found in the literature (e.g. \cite{liu2019deformation, lallemant1980rheology}), but these rocks are a mixture of olivine, pyroxene and other secondary phases and their microstructural evolution may be controlled by their polymineralic nature \cite{Furstoss2020} in addition to the impurity effect. However, in natural polymineralic peridotites like harzburgites or lherzolites, the maximum grain size is dictated by the spacing between static, pinning phases like spinels, and is always larger than their actual GS \cite{Herwegh2011}. An alternative could be to focus on dunites, a coarse-grained rock, which mineral assemblage is greater than 90\% olivine with only minor amounts of secondary minerals. However, in such rocks, the initial crystal size distribution is clearly related to the mechanism of their formation, which usually involves extensive interaction with melts percolating the mantle and may be very different from the GS distribution used as initial conditions for our models \cite{berger1984dunites}, \cite{kelemen1990reaction}. Considering these two limitations, we chose to compare our models results with classical (i.e., polymineralic) peridotites, being aware that our model only captures a part of the processes that limit the maximum grain size of natural rocks.

\noindent
Considering only rock samples with textures typical of thermal annealing and according to the terminology defined by \cite{harte1977rock} we selected samples which were described as “coarse”, which is equivalent to “protogranular” \cite{mercier1975textures}, and equant (or granular). Porphyroclastic, granuloblastic and tabular textures were not considered because they reflect significant rock deformation that was not subsequently fully annealed by grain growth, or grain growth in presence of melts or fluids. We briefly recall here the geological setting, age and temperature history of the selected samples.

\noindent
\textbf{The Udachnaya (Siberia) kimberlite xenoliths}

The Udachnaya kimberlite pipe in Siberia is well known due to its diamond mine and because of the occurrence of megacrystalline olivine ($\approx$ 10cm grain size) in harzburgitic and dunitic xenoliths \cite{pokhilenko2014new}. The age of Udachnaya pipe has been determined between 345 and 385 Ma depending on the dating method \cite{dehvis1980new, ilupin1990kimberlites}. Temperature estimates for the megacrystalline peridotite xenoliths range between 1173 and 1373K \cite{pokhilenko1993megacrystalline, griffin1996thermal} for a depth estimated around 150-200 km in the thick lithosphere of the Siberian Craton. While the temperature range and grain size of these peculiar rocks are well constrained, it is barely impossible to determine precisely how much time they have spent at these temperatures before being erupted in Late Devonian – Early Carboniferous times. Re/Os ages in diamond sulfides have provided an age for the formation of the cratonic lithosphere around  1.8 Ga \cite{ionov2015age}, so, if we assume that these rocks represent samples from the oldest part of the cratonic lithosphere, we can estimate the maximum residence time of these rocks at 1173-1373K at 1800-350 = 1450 Ma. A minimum residence time is difficult to estimate, since the cratonic mantle has been modified by metasomatism (interaction with percolating melts) after its stabilization \cite{ionov2015age}.

\noindent
\textbf{The Kaapvaal (South Africa) kimberlite xenoliths}

The Archean lithosphere of the Kaapvaal craton in South Africa stabilized around 3 Ga ago according to Re-Os isotope studies \cite{pearson1995re}. Mantle xenoliths have been sampled by magmatism in Late Jurassic to Cretaceous times, i.e. between 180 and 90 Ma \cite{griffin2014emplacement}. Most peridotite xenoliths consist of coarse (5-8 mm) to cm-size grained harzburgites indicating significant annealing posterior to an early stage of deformation \cite{baptiste2014petrophysical}. PT estimates range along a low geothermal gradient with temperatures of $873-1273K$ at depths between 80 and 150 km \cite{chu2012, saltzer2001spatial, baptiste2014petrophysical}. In the Jagersfontein pipe, very coarse-grained peridotite xenoliths exhibit grain sizes from 5 to 20 mm with temperature estimates ranging from $973-1223K$ \cite{winterburn1990peridotite}. Olivine CaO content reaches up to 1200 ppm \cite{hervig1986lherzolite}. For our model, we assume that a mean temperature of 1073K was reached rapidly after the stabilization of the cratonic lithosphere, and we consider that these rocks have spent 2 to 3 Ga at this temperature before being erupted.

\noindent
\textbf{The Kerguelen hotspot in the Indian Ocean}

The Kerguelen archipelago is part of a Large Igneous Province, the Kerguelen Plateau, formed above the Kerguelen plume \cite{bascou2008integrated, mattielli1996kerguelen, gregoire1995kerguelen}. Plume-related volcanism forming the Kerguelen Islands started around 45 Ma ago and lasted until 0.1 Ma ago (\cite{cottin2011kerguelen} and references therein). Ultramafic xenoliths brought at the surface in the Kerguelen Islands by the plume-related volcanism are harzburgites and dunites typical of a depleted mantle which has undergone a large degree of partial melting (\cite{bascou2008integrated} and references therein). Equilibrium PT conditions determined on xenoliths close to the crust-mantle boundary are around 1 GPa and 1173-1273K \cite{gregoire1995kerguelen}. Some protogranular harzburgites have mean grain sizes of 2-10 mm while equigranular dunites have a mean grain size between 0.5 and 1 mm \cite{bascou2008integrated}. Here again, it is difficult to estimate the annealing time of these xenoliths. Based on geochemical and petrological analyses the Kerguelen harzburgites were interpreted as residues from a partial melting episode linked with the Kerguelen plume, that were subsequently affected by melt percolation forming the dunites \cite{mattielli1996kerguelen, bascou2008integrated}. Peridotite xenoliths have been sampled in lavas dated between 28 and 7 Ma (\cite{bascou2008integrated} and references therein) so they might have spent up to 38 Ma at temperatures close to 1173-1273K.

\noindent
The temperature, rock type and mean grain size for different contexts raise above are summarized within table \ref{tab1}.

\begin{table}[H]
	\begin{center}
		\hbox to \hsize{\hss\includegraphics*[width=\textwidth]{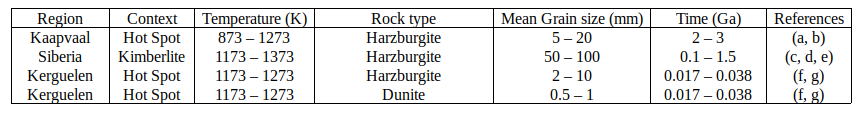}\hss}
		\caption{Summary of constraints on temperature, grain size and geodynamic context for the four selected peridotite samples that were used to estimate our model performance. $a$ \protect\cite{baptiste2014petrophysical}; $b$ \protect\cite{winterburn1990peridotite}; $c$ \protect\cite{pokhilenko2014new}; $d$ \protect\cite{goncharov2012thermal}; $e$ \protect\cite{yudin2014age}; $f$ \protect\cite{gregoire1995kerguelen}; $g$ \protect\cite{bascou2008integrated}.}
		\label{tab1}
	\end{center}
\end{table}

The geological contexts presented above will be used in the following to test the performance of our mean-field solute drag model. In a first step, in order to calibrate the solute drag parameters on natural constraints for the full-field model, we will consider that the model should produce grain sizes ranging between 0.5 and 10mm for annealing times ranging between 0.1Ma and 1Ga. For the full field models being supposed to fit those natural constraints, we took a temperature of 1073K because this temperature corresponds to a typical upper mantle temperature. Moreover, it corresponds to the temperature at which diffusion becomes so slow that textural rebalancing under static conditions becomes negligible.

\section{Results}

In this section, we present the full field and mean field results concerning the long term annealing of olivine aggregates. The olivine grain boundary energy ($\gamma$), calcium concentration within grain matrix ($c_0$) are taken as $1J.mm^{-1}$ \cite{CooperAndKohlstedt} and $100ppm$ respectively while the grain boundary mobility ($M$) is taken as an Arrhenius's law where the reference value and activation energy are equal to $M_0 = 4.10^4 mm^4.J^{-1}.s^{-1}$, $Q_M = 185kJ.mol^{-1}$ \cite{Furstoss2020} respectively.

\noindent
To perform long term annealing full field simulations, passing from micrometer to millimeter scale grain sizes, we need to define a model chaining strategy. The full field simulations begin with approximately 2000 grains respecting an initial grain size distribution corresponding to the one used in laboratory experiments \cite{Karato89}. When the number of grains within the simulation domain is less than 200, the simulation is stopped and a new set of 2000 grains and a larger domain is generated for the next job following the final grain size distribution of the latest run. The error due to this chained calculation is minimized by sampling very precisely the final GSD and by imposing it as the new initial GSD using the Vorono\"\i-Laguerre Dense Sphere Packing algorithm \cite{hitti2012precise}.

\noindent
In the following, we first propose an adjustment of the solute drag material parameters, based on full field simulations, which permits to reconcile laboratory and natural observations by producing the adequate GG kinetics. Then, the mean field model presented within section \ref{MeanField} is calibrated on full field results. Finally we show an application of this framework within the geological contexts presented in section \ref{BiblioGeol}.

\subsection{Full field and mean field simulations : adjustment of solute drag parameters}

\begin{figure}[H]
	\begin{center}
		\hbox to \hsize{\hss\includegraphics*[width=\textwidth]{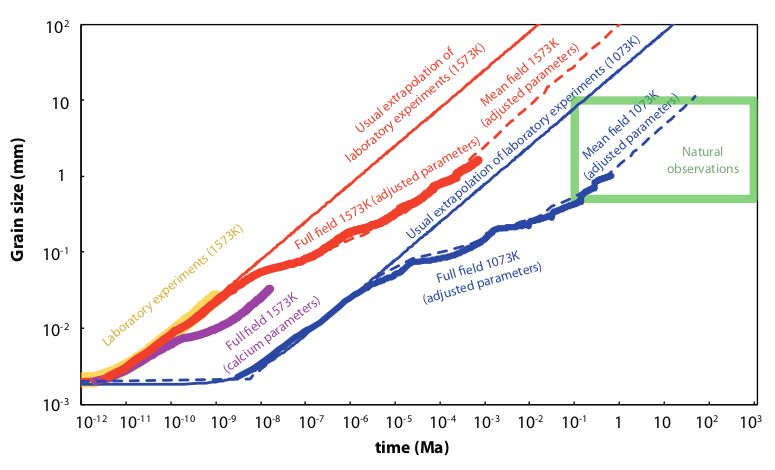}\hss}
		\caption{Olivine aggregates GG kinetics at 1573K and 1073K. Thin solid lines represent the usual extrapolations (not accounting for solute drag) at 1573K (red) and 1073K (blue) of the laboratory results \cite{Karato89} (yellow). The bold purple line present the predictions from full field simulation using calcium solute drag parameters $\alpha_{Ca}$ and $\beta_{Ca}$ calculated in section \ref{SoluteDragOl}, while the bold red and blue solid lines represent full field simulations with $\alpha = \alpha_{Ca}$ and $\beta = 100 \beta_{Ca}$ at 1573K and 1073K, respectively . The dashed lines represent the GG kinetics computed using the mean field model with adjusted parameters based on the full field simulations at 1573K (red) and 1073K (blue). The green rectangle represents olivine mean grain sizes and annealing times typical of lithospheric mantle rocks (see section \ref{BiblioGeol}).}
		\label{fig2}
	\end{center}
\end{figure}

We begin our full field simulations using the solute drag parameters $\alpha_{Ca}$ and $\beta_{Ca}$ presented in section \ref{SoluteDragOl}. However at 1573K the computed GG kinetics does not match the experimental results \cite{Karato89} (Fig.\ref{fig2}). The GG curve quickly deviates from the laboratory data and thus we stopped this simulation at the second run.

\begin{figure}[H]
	\begin{center}
		\hbox to \hsize{\hss\includegraphics*[scale=0.2]{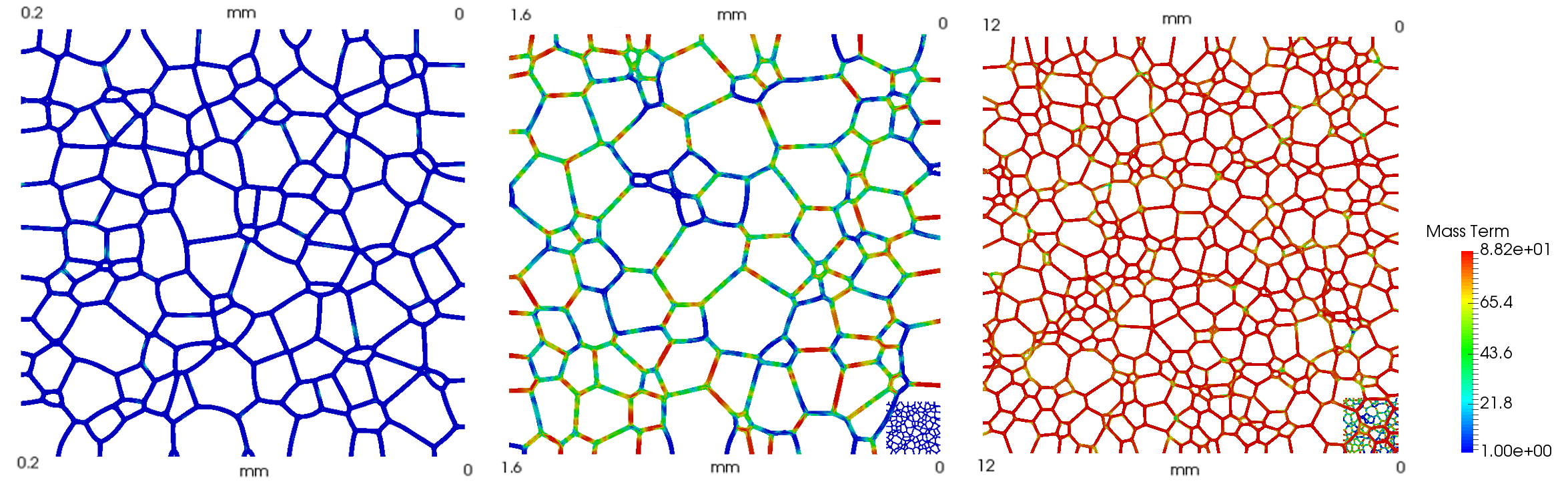}\hss}
		\caption{Distribution of mass term $\mathcal{M}$ along grain boundaries for the first three runs at 1573K using $\alpha = 10 \alpha_{Ca}$ and $\beta = 500 \beta_{Ca}$. The first run (left), a $0.2mm$ aside square, is also represented in the right bottom corner of the second run (center), a $1.6mm$ aside square, which is also represented at the right bottom corner of the third run (right), a $12mm$ aside square.}
		\label{fig3}
	\end{center}
\end{figure}

We tested different values for the solute drag parameters and found that $\alpha = \alpha_{Ca}$ and $\beta = 100 \beta_{Ca}$ produce results consistent with laboratory results at 1573K and have GG kinetics compatible with natural observations at 1073K (Fig.\ref{fig2}). Using those values, the mean grain size evolution begins to deviate from the classical extrapolation of laboratory results (which does not account for solute drag) for a mean grain size near 30$\mu m$ at 1573K and near 50$\mu m$ at 1073K (Fig.\ref{fig2}). The GG kinetics is slowed down by solute drag for mean grain sizes of up to ca. 1mm. For coarser mean grain sizes, GG is weakly influenced by solute drag (low velocity regime in Fig.\ref{fig0}). The distribution of the mass term $\mathcal{M}$ (see section \ref{LSImpl}, Eq.\ref{eqMassTerm}) within the simulated microstructure (Fig.\ref{fig3}) clearly shows an increase of its value from the small grains to the large ones.

At first, for ultra-fine grained aggregates as the ones used in laboratory experiments, the mass term is nearly equal to 1 within all the microstructure (Fig.\ref{fig3}). This first phase correspond to the high velocity regime in which the grain growth kinetics is nearly unimpacted by the solute drag (Fig.\ref{fig0}). Afterwards, as the grain sizes increase the mass term becomes heterogeneously distributed between one and its maximum value depending on the local grain boundary velocities, which are controlled by the local curvature. Within this regime, the long straight grain boundary segments have an higher mass term and their velocities are near the velocity the most impacted by solute drag. Those segments are thus slowed down by solute drag. They also exert a pinning force on other grain boundaries, which enhances the deceleration of the grain growth kinetics. Finally, when grain sizes are sufficiently large, the grain boundary velocities become very small and one enters the low velocity regime in which the grain growth kinetics is again weakly impacted by the solute drag (Fig.\ref{fig0}). The mass term almost reaches its maximum value in the whole microstructure (Fig.\ref{fig3}).

\begin{figure}[H]
	\begin{center}
		\hbox to \hsize{\hss\includegraphics*[width=\textwidth]{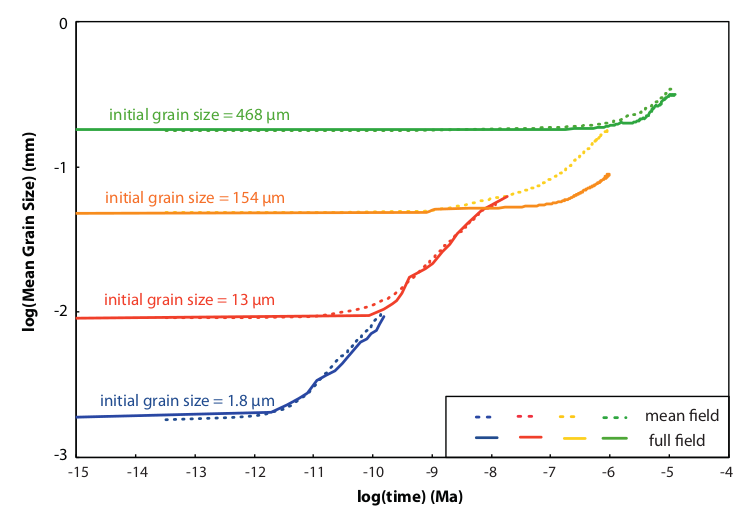}\hss}
		\caption{Mean grain size evolution at 1573K for $\alpha = 10 \alpha_{Ca}$ and $\beta = 500 \beta_{Ca}$ for the first four runs (without applying the time offset), the full lines represent the full field model and the dashed lines represent the mean field computations.}
		\label{fig4}
	\end{center}
\end{figure}

\noindent
The analysis of the GG kinetics for simulations with different initial mean grain sizes highlights a first phase of very slow or even null mean grain size evolution (Fig.\ref{fig4}).

\begin{figure*}[ht!]
	\begin{center}
	\begin{subfigure}[h]{0.5\textwidth} 
		\includegraphics*[width=\columnwidth]{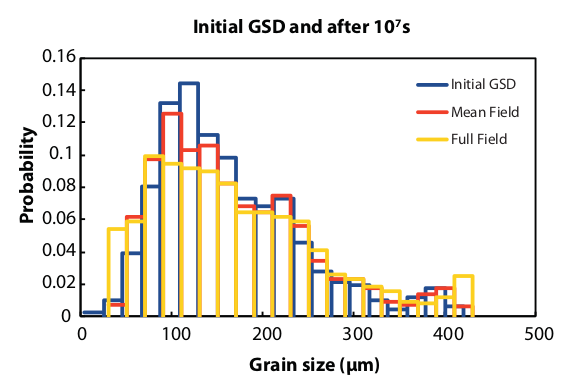}
		\caption{}
		\label{Fig7_a}
	\end{subfigure}%
	\begin{subfigure}[h]{0.5\textwidth} 
		\includegraphics*[width=\columnwidth]{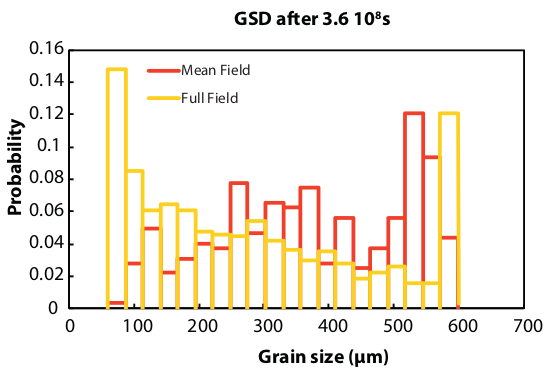}
		\caption{}
		\label{Fig7_b}
	\end{subfigure}%
  \caption{GSD (in number) for the fourth run presented in figure \ref{fig4} (at 1573K for $\alpha = 10 \alpha_{Ca}$ and $\beta = 500 \beta_{Ca}$), \ref{Fig7_a} : initial GSD (blue), mean field (orange) and full field (yellow) modeled GSD after $10^7s$, \ref{Fig7_b} : mean field (orange) and full field (yellow) modeled GSD after $3.6.10^8s$.}
  \label{figGSD}
	\end{center}
\end{figure*}

\noindent
The duration of this phase of slow GG is longer when the initial mean grain size is coarse (Fig.\ref{fig4}). This initial slow GG kinetics can be explained by analyzing the GSD (Fig.\ref{figGSD}). In fact, even when the initial microstructure is composed of coarse grains, some grains have to shrink to let the other ones grow. This can be seen by comparing in Fig.\ref{Fig7_a} the GSD of the initial microstructure and at the end of the initial slow GG phase for the experiment with the coarser initial mean grain size in Fig.\ref{fig4}. This comparison highlights an enrichment in small grains (Fig.\ref{Fig7_a}). Before disappearing, the GBM velocities of the shrinking grains will necessarily pass through a velocity regime highly impacted by the presence of impurities. This will affect the grain growth kinetics directly (by slowing down the shrinking) and indirectly by impeding the movements of the other grain boundaries through pinning mechanisms.

\noindent
From these observations, we infer that initial grain size does not affect GG kinetics apart from delaying the start of the rapid GG stage.

\noindent
After calibration, the mean field model (Eq.\ref{HillertModifiedMF}) reproduces well the full field modeled GG kinetics for both temperatures (Figs.\ref{fig2} and \ref{fig4}). The best fitting mean field parameters $C_\alpha$, $C_\beta$ and $n$ are equal to $0.7$, $0.25$ and $0.15$ respectively. The GSD predicted by the mean field approach is consistent with the full field modeled ones (Fig.\ref{figGSD}) even if some differences can be seen after some annealing time (Fig.\ref{Fig7_b}). This mean field approach allows to be predictive on olivine aggregates GG kinetics for a much lower computational cost, which permits to test our formalism on different geological contexts.

\subsection{Implication for the microstructural evolutions in ultramafic rocks}

In order to estimate how well our mean field model predicts the average grain size of natural samples with reasonable $c_0$ (Ca concentration within bulk) values, we test it against different contexts in terms of annealing temperature and measured grain sizes.

\subsubsection{Mean field GG models} \label{HalfSpaceCooling}

For the mean field GG model, the temperature is kept constant and grain size grows indefinitely; we compute the GG curve for residence times up to 2 Ga. GG curves are computed from four temperatures of 1073, 1173, 1273 and 1373K and $c_0$ values of 600, 800, 1000 and 1200 ppm.

\begin{figure}[H]
		\begin{center}
		\hbox to \hsize{\hss\includegraphics*[width=\textwidth]{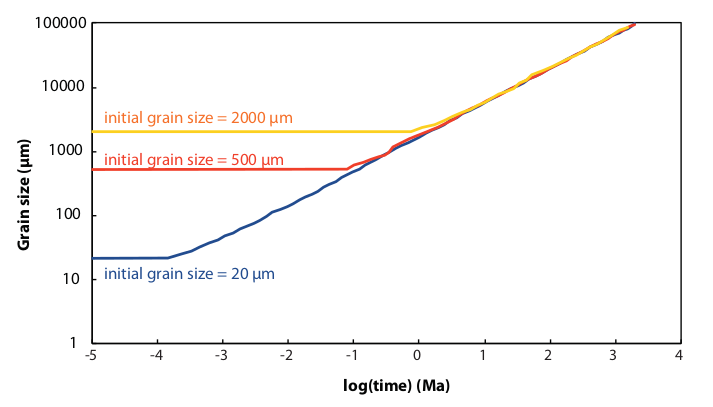}\hss}
		  \caption{Grain size evolution for isothermal model with $T=1173K$, $c_0 =800ppm$ and three initial grain sizes of 20, 500 and 2000$\mu m$.}
		  \label{figIndepGSInit}
	\end{center}
\end{figure}

\noindent
All models were run initially with an initial grain size of 20$\mu m$ and a standard deviation of 2$\mu m$; then, isothermal models were run again with an initial grain size of 0.5mm and a standard deviation of 50$\mu m$ and an intial grain size of 2mm and a standard deviation of 200$\mu m$. In fact, the initial grain size has no effect of the grain growth curve or final grain size, apart from shifting the beginning of the beginning of the positive slope on the grain growth curve (Fig.\ref{figIndepGSInit}).

\subsubsection{Results}

\begin{figure}[H]
	\begin{center}
		\hbox to \hsize{\hss\includegraphics*[width=\textwidth]{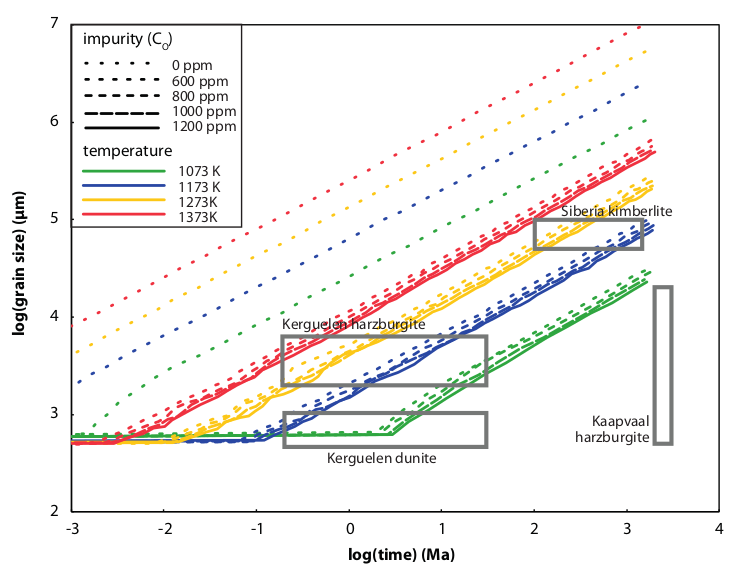}\hss}
		\caption{Grain size evolution for isothermal models with 4 different temperatures and 4 different impurity ($c_0$) concentrations. Solid squares indicate the approximate location of Siberia and Kerguelen and Kaapvaal xenoliths on this curve according to the literature (see references in Tab.\protect\ref{tab1}).}
		\label{fig6}
	\end{center}
\end{figure}

The $c_0$ parameter, in the range of tested values, has only a moderate effect of the final or intermediate grain sizes (table \ref{tab2} and figure \ref{fig6}). However, these grain sizes are considerably smaller than the ones predicted without impurities. Indeed, for isothermal models, grain sizes reach 2 to 8 meters after 1 Ga at temperatures between 1173 and 1373K, respectively, while they do not exceed 47 cm for the scenario with the largest temperature (1373K) and the lowest impurity concentration (600 ppm).

\begin{table}[H]
	\begin{center}
		\hbox to \hsize{\hss\includegraphics*[width=\textwidth]{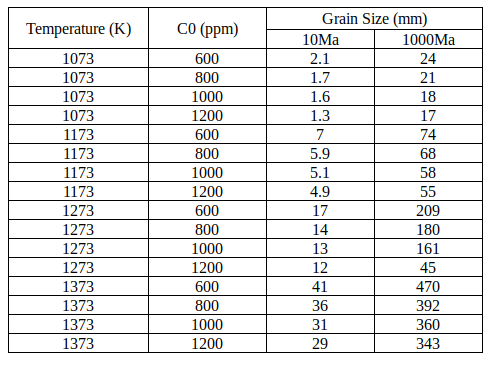}\hss}
		\caption{Temperature and geochemical ($c_0$) conditions for mean field models and their results in terms of grain sizes at two different steps (10 and 1000 Ma).}
		\label{tab2}
	\end{center}
\end{table}

\noindent
Mean field models predict a GG rate which decreases exponentially with time (Fig.\ref{fig6}). Extremely large grain sizes of several cm can be reached for conditions (temperature and annealing times) consistent with those inferred for the Udachnaya kimberlite xenoliths (i.e., temperatures of 1173 to 1273K and residence times between 100 and 1450 Ma). The Kaapvaal peridotites, which have smaller grain sizes and potentially larger residence times than the Udachnaya ones, range slightly below the 1073K temperature curve.

\noindent
For annealing times of 1 to 40 Ma, predicted mean grain sizes range between 2 mm (for 1173K and $c_0$ = 1200 ppm) and 4 cm (for 1273K and $c_0$ = 600 ppm). These predictions match the grain sizes of the harzburgites from the Kerguelen Islands. The grain size measured in Kerguelen dunites is smaller than these predictions; it suggests either rapid exhumation (hence very limited grain growth) or a lower residence temperature ($\approx$ 1073K).

\section{Discussion}

The GG kinetics simulated using $\beta_{Ca}$ cannot adequately reproduce both laboratory experiments and natural observations on grain size evolution of olivine aggregates (Fig.\ref{fig2}). However we find that a value of $\beta$ two orders of magnitude larger than $\beta_{Ca}$ allows our models to be consistent with those two constraints. This difference could arise from two main reasons :

\noindent
First, the two functions $D(x)$ and $E(x)$ needed for the calculation of solute drag parameters $\alpha$ and $\beta$ (see section \ref{SoluteDragTh}) are poorly known even for metallic materials in which the interaction between an impurity ion and a grain boundary is mostly elastic. For ceramics-like materials such as olivine, this interaction should also account for electrostatic interaction between the solute ion and the grain boundary and the interaction between solute-vacancy dipoles and electrical field around grain boundaries \cite{yan1983space}. The quantification of all of those interactions should be done by atomistic calculations using systematic approaches for describing grain boundaries and their interactions with solute \cite{hirel2019systematic}.

\noindent
Secondly, it is well accepted that the impurity concentration at grain boundary evolves with the grain size, increasing when grain size increases \cite{marquardt2018structure} which could also be expressed by a grain size dependent partition coefficient \cite{hiraga2004seggregation}. Even if the solute drag pressure (Eq.\ref{EqImpurityDragPressure}) depends on grain matrix impurity concentration, one can find expressions for $\alpha$ and $\beta$ as functions of the partition coefficient \cite{cahn1962impurity}. The expected effect of indirectly introducing this grain boundary impurity concentration may be similar to an increase of $\beta$, as we did here. In fact, increasing the $\beta$ will increase the space of the high velocity regime for which the grain boundary migration will be poorly affected by impurities. An increase of the impurity concentration with grain size will also result in a lower impact of the solute drag for small grains.

Our full field and mean field simulations show an initial phase of very slow grain size evolution (Figs.\ref{fig4} and \ref{figIndepGSInit}) due to the direct slowed down of the grain shrinkage and growth by solute drag, and indirectly by the impediment of GBM by the impurity slowed grain boundaries. This quasi-static phase, which is longer for coarser initial grain sizes, implies a very weak dependency of the grain growth kinetics on initial grain size. This small dependency  frees us from the need of precise constrains on the initial grain size, which is very difficult to know in geological contexts. Taking advantage of this, we can apply our solute drag model for different stable geological contexts. For the majority of the geological contexts presented here, our formalism shows consistent grain size / time predictions (Fig.\ref{fig6}). It is difficult to evaluate precisely the performance of our models with respect to the evolution of natural samples, given the lack of data concerning their temperature and grain size evolution. However, impurity drag due to Ca concentration in olivine, in the range of commonly observed value, explains a grain size reduction of several orders of magnitude at geological timescales, compared to models without impurities.

\section{Conclusions}

In this work we have demonstrated that accounting for the presence of impurities within olivine rich-rocks permits to explain GG kinetics of both experimentally and naturally annealed olivine aggregates. The solute drag parameters needed for this approach are however quite different from the ones expected for the calcium, which is the mantle rock impurity expected to be the most impacting on olivine grain boundary migration. Atomistic calculations or a strong experimental framework could be needed to explain this gap.

\noindent
We have developed a new mean field model accounting for the presence of those impurities and showed this approach could be used to predict grain size of olivine-rich rocks in geological contexts such as oceanic cooling or isothermal evolution. As this approach successfully reproduced natural grain size in annealed peridotite, healing kinetics may be implemented in large scale numerical geodynamic models based on this framework. In order to have a useful grain size evolution law for geodynamic model, a model accounting for the competition between grain size reduction due to deformation and growth should be developed.

\noindent
In order to be consistent with the real multiphase nature of natural rocks, the influence of second phases, such as pyroxenes and alumina phases, on GG kinetics should also be considered. Such a framework may open the door to a paleo-chronometer based on grain size evolution within geological contexts in which deformation has stopped and GG predominates.

\section{Acknowledgments}

\noindent
This work was supported by CNRS INSU 2018-programme TelluS-SYSTER. 

\noindent
The support of the French Agence Nationale de la Recherche (ANR), ArcelorMittal, FRAMATOME, ASCOMETAL, AUBERT\&DUVAL, CEA, SAFRAN through the DIGIMU Industrial Chair and consortium are gratefully acknowledged.

\section{Data availability}

The data for supporting all figures of the paper are avalaible upon request to the authors as well as the mean field code used in this work.

\bibliographystyle{unsrt}
\bibliography{thebibliography}
\end{document}